# C-MADA: Unsupervised Cross-Modality Adversarial Domain Adaptation framework for medical Image Segmentation

Maria Baldeón Calisto[1][0000-0001-9379-8151] and Susana K. Lai-Yuen[2][1111-2222-3333-4444]

[1] Universidad San Francisco de Quito, Diego de Robles s/n y Vía Interoceánica, Quito, Ecuador
[2] University of South Florida, Tampa, FL, USA
mbaldeonc@usfq.edu.ec

**Abstract.** Deep learning models have obtained state-of-the-art results for medical image analysis. However, when these models are tested on an unseen domain there is a significant performance degradation. In this work, we present an unsupervised Cross-Modality Adversarial Domain Adaptation (C-MADA) framework for medical image segmentation. C-MADA implements an image- and feature-level adaptation method in a sequential manner. First, images from the source domain are translated to the target domain through an unpaired image-to-image adversarial translation with cycle-consistency loss. Then, a U-Net network is trained with the mapped source domain images and target domain images in an adversarial manner to learn domain-invariant feature representations. Furthermore, to improve the network´s segmentation performance, information about the shape, texture, and contour of the predicted segmentation is included during the adversarial training. C-MADA is tested on the task of brain MRI segmentation, obtaining competitive results.

## 1 Introduction

In the last decade, deep convolutional neural networks (CNNs) have become the state-of-the-art models for automatically segmenting medical images [1]. However, CNNs require a massive amount of labelled data to achieve a high performance and assume that the training/source and test/target dataset follow the same probability distribution. Hence, when tested on images that pertain to a different distribution, their performance degrades in proportion to the distribution difference.

Recently, unsupervised domain adaptation (UDA) techniques have gained attention to reduce domain shift. Most UDA models can be broadly categorized into feature-level adaptation methods [2], image-level adaptation methods [3], and combined image- and feature-level adaptation methods [4]. The latter technique has proven to provide a better segmentation performance as both perspectives are complementary.

In this work, we present C-MADA, an unsupervised **C**ross-**M**odality **A**dversarial **D**omain **A**daptation framework for medical image segmentation. C-MADA implements an image- and feature-level adaptation method in a sequential manner. First, images from the source domain are translated to the target domain through an unpaired image-to-image adversarial translation [5]. Then, a U-Net network is trained with the source and target domain images in an adversarial manner to learn domain-invariant feature representations and produce probable segmentations for the target domain. Furthermore, to improve the network´s segmentation performance, information about the shape, texture, and contour of the predicted segmentation is included during the adversarial training. C-MADA is evaluated on the problem of tumor and chochlea segmentation from brain MRIs from the crossMoDa Grand Challenge. Our method achieves a competitive performance on the validation phase.

## 2   Methods

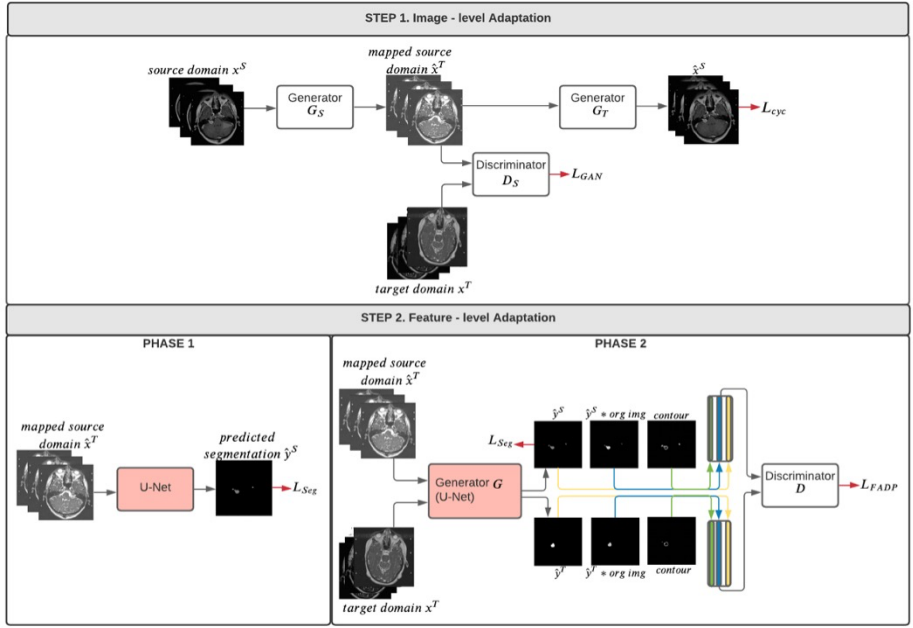

**Fig. 1.** Two-step C-MADA framework. Step 1 performs and image-level adaptation. Step 2 implements a feature-level adaptation.

C-MADA is composed of two sequential steps: the image-level adaptation and the feature-level adaptation as presented in Fig. 1. These steps are explained next.

### 2.1   Image-level Adaptation

The training dataset is comprised of labeled images $\{(x^s, y^s)\}$ from the source domain $S$, and unlabeled images $\{x^T\}$ from the target domain $T$. In this step images from $S$ are mapped to $T$ in terms of appearance by implementing CycleGAN [5]. In this model, a generator network transforms the source domain images to target domain $G_S: x^S \to \hat{x}^T$. Meanwhile, a discriminator network $D_S$ aims to distinguish $x^t$ from $\hat{x}^T$. Hence, $G_s$ and $D_s$ compete in a two-player minimax game where $G_s$ aims to minimize the objective function displayed in Eq. 1, while $D_s$ tries to maximize it.

$$\mathcal{L}_{GAN}(G_S, D_S) = \mathbb{E}_{x^t \sim T}[log D_s(x^t)] + \mathbb{E}_{x^s \sim S}[log(1 - D_s(G_s(x^s)))] \quad (1)$$

Furthermore, to guarantee the information in $x^s$ is preserved during the transformation, another mapping function from target domain to source domain $G_T: x^T \to \hat{x}^S$ is implemented. Likewise, a discriminator $D_T$ aims to discriminate between $x^s$ and $\hat{x}^S$. $G_T$ and $D_T$ are trained with a similar loss as in Eq. 1. Moreover, to incentivize the translation to be cycle consistent, the cycle consistent loss displayed in Equation 2 is added to the objective function in Equation 1.

$$\mathcal{L}_{cyc}(G_S, G_T) = \mathbb{E}_{x^s \sim S}[\|G_T(G_s(x^S)) - x^S\|] + \mathbb{E}_{x^t \sim T}[\|G_s(G_T(x^T)) - x^T\|] \quad (2)$$

The full objective function being optimized during this step is presented in Eq. 3. The generator networks have a ResNet architecture [6], while the discriminators have a PatchGAN architecture [7]. Moreover, an $\lambda=10$ is used.

$$\mathcal{L}_{adv}(G_S, G_T, D_S, D_T) = \mathcal{L}_{GAN}(G_S, D_S) + \mathcal{L}_{GAN}(G_T, D_T) + \lambda \mathcal{L}_{cyc}(G_S, G_T) \quad (3)$$

### 2.2 Feature-level Adaptation

In step 2, a 2D U-Net architecture is trained to segment the target domain images in two phases. In the first phase, the U-Net is fully trained in a supervised manner using the translated source dataset. The loss function being minimized is a linear combination of the Dice coefficient and cross-entropy loss as presented in Eq. 4, where $y_{ic}$ and $\hat{y}_{ic}$ are the ground-truth label and the predicted probability for pixel $i$ in class $c$, respectively. $\alpha_c$ are weight parameters for the dice loss in class $c$, and $\beta$ a weight parameter for the dice loss. We set $\alpha_0 = .1$, $\alpha_1 = .4$, $\alpha_2 = .5$, and $\beta = 0.65$.

$$\mathcal{L}_{seg} = \beta \sum_c \alpha_c \left(1 - \frac{2 \sum_i \hat{y}_{ic} y_{ic}}{\sum_i \hat{y}_{ic} + \sum_i y_{ic}}\right) + (1-\beta) \sum_c \sum_i (y_{ic} \log(\hat{y}_{ic}) + (1 - y_{ic}) \log(1 - \hat{y}_{ic})) \quad (4)$$

In the second phase, a feature-level adaptation method is applied through an adversarial learning scheme. As shown in Fig. 1, the fully trained U-Net takes the role of the generator $G$ and predicts the segmentations for $\hat{x}^T$ and $x^T$ as $G(\hat{x}^T) = \hat{y}^S$ and $G(x^T) = \hat{y}^T$, respectively. A discriminator $D$ receives as input these predicted segmentations, in addition to boundary information, and aims to discriminate $\hat{y}^S$ from $\hat{y}^T$. Meanwhile, $G$ is trained to trick the discriminator by producing probable target domain segmentations. The objective function being optimized in this phase is presented in Eq. 5.

$$\mathcal{L}_{FADP}(G, D) = \mathbb{E}_{x^t \sim T}[log D_s(\hat{y}^S)] + \mathbb{E}_{x^s \sim S}[log(1 - D_s(\hat{y}^T))] \quad (5)$$

Inspired by [2], the input to $D$ is the concatenation of the predicted segmentation, the elementwise multiplication of the predicted segmentation and original image, and the contour of the predicted segmentation by applying a Sobel operator. This input provides $D$ with enough information about the shape, texture, and contour of the segmented region to force the U-Net to be boundary and semantic-aware. Finally, to prevent catastrophic forgetting from the source domain, during each training iteration the U-Net is trained in a supervised manner with the mapped source domain images and using the loss function in Eq. 4.

For selecting the network´s hyperparameters and weights for testing, we propose a validation loss function based on an area ratio metric. Using the ground truth segmentations from the source domain, we compute the average number of pixels that are part of class $c$ per slice ($SAvgPix_c$). We assume $SAvgPix_c$ is a good approximator of the average number of class $c$ pixels in the target domain segmentation. Therefore, we compute the average number of pixels predicted to be part of class $c$ in the target domain segmentation and divide it by $SAvgPix_c$. We prefer configurations whose values are close to 1. Moreover, to include information about the segmentation performance, to the loss function we add the dice loss in each class on the source domain. Hence the configuration that minimizes this loss function is selected for testing.

## 3 Results

C-MADA is evaluated on the task of brain MR image segmentation [8, 9]. Source and target domain images are resampled to a spatial resolution of 0.468 × 0.468 × 1.5mm and set to a fixed size of 448 × 448 × 120 voxels. Moreover, all pixel intensities are clipped within the 3 standard deviations of the mean. The adversarial model in the image-level adaptation is trained for 40 epochs. Meanwhile, the U-Net is trained for 500 epochs during phase 1 and the adversarial model of phase 2 for 100 epochs. The model was implemented in pytorch and trained using a 32 GB V100 GPU.

In Table 1 we present the mean dice coefficient and mean average symmetric surface distance (ASSD) for the tested models. In Table 1, S1+ *Network* refers to the implementation of step 1 of the framework and then fully training the *Network* with the mapped source domain images. On the other hand, C-MADA ([seg]) refers to implementing step 1 and 2 of the proposed framework but the input to the discriminator $D$ is only the predicted segmentation. The results demonstrate that each step of the proposed framework improves the segmentation performance. We also tested training the U-Net directly on the source domain dataset, applying only Phase 2 of the framework, varying the input to discriminator D and the segmentation network for phase 2. Nevertheless, the proposed validation loss function was low in those configurations and decided not to submit the results to the challenge.

Table 1. Results of C-MADA framework on the CrossMoDa challenge validation set

| Method | VS Dice | VS ASSD | Cochlea Dice | Cochlea ASSD |
| --- | --- | --- | --- | --- |
| **C-MADA** | **0.646±12.067** | **7.256±0.162** | **0.401±0.162** | **3.672±9.631** |
| C-MADA([seg]) | 0.597±0.281 | 9.679±15.346 | 0.384±0.176 | 1.932±4.055 |
| S1+residualU-Net | 0.594±0.280 | 7.144±8.286 | 0.361±0.143 | 5.413±6.667 |
| S1 + U-Net | 0.582±0.285 | 9.778±16.831 | 0.265±0.146 | 2.174±4.034 |